\documentclass[12pt]{article}
\usepackage{amsmath}
\usepackage{amssymb}
\usepackage{myart}

\normalbaselines
\input tcilatex
\begin{document}

\title{Incompleteness of the particle dynamics in microcosm and the skeleton
conception of elementary particles as overcoming of this incompleteness }
\author{Yuri A.Rylov}
\date{Institute for Problems in Mechanics, Russian Academy of Sciences,\\
101-1, Vernadskii Ave., Moscow, 119526, Russia.\\
e-mail: rylov@ipmnet.ru\\
Web site: {$http://rsfq1.physics.sunysb.edu/\symbol{126}rylov/yrylov.htm$}\\
or mirror Web site: {$http://gasdyn-ipm.ipmnet.ru/\symbol{126}%
rylov/yrylov.htm$}}
\maketitle

\begin{abstract}
The skeleton conception of elementary particles is considered in the paper.
Conventional particle dynamics is formulated in an unaccomplished form,
which is adequate only in the continuous space-time geometry. The
conventional differential equations of the particle motion cannot be written
in the discrete space-time geometry. In the discrete space-time geometry the
particle world line is replaced by the world chain. The world chain links
has a finite length (not infinitesimal). The world chain appears to be
stochastic. Statistical description of stochastic world chains leads to the
Schr\"{o}dinger equation, if the elementary length of the geometry is chosen
in a proper way. The quantum principles are founded by existence of the
discrete (and multivariant) space-time geometry and lose the role of prime
physical principles. In the skeleton conception the particle is described by
its skeleton (several rigidly connected space-time points). Skeleton
conception of the elementary particles realizes a proper description of the
particle state, which appears to be adequate in a discrete space-time
geometry. The particle dynamics takes the form of a monistic conception,
which is described completely in terms of the world function of the
space-time geometry. The skeleton conception accomplishes the transition
from nonrelativistic physics to the relativistic one and realizes the total
geometrization of particle dynamics.
\end{abstract}

\section{Introduction}

A conception of the elementary particles is the first stage of the
elementary particle theory. The conception considers concepts of the theory,
logical connection between the concepts, their compatibility between
themselves and with the relativity principles. Some concrete statements of
the conception (for instance, a choice of the space-time geometry) are not
formulated. As a result the conception of elementary particles cannot be
tested experimentally, because it cannot make any concrete predictions. At
the next step of the theory development, when a concrete space-time geometry
is chosen, the theory can make predictions, which can be tested
experimentally. Consideration of a conception of elementary particles is
interesting in the relation, that it may show, \textit{which concepts cannot
be used in the elementary particle theory}.

Now, hundred years after construction of the special relativity the
statement, that the special relativity theory is not yet accomplished, looks
very strange and unexpected. However, transition from nonrelativistic
physics to the relativistic one concerns mainly dynamic equations,
describing particle motion. Concept of the particle state remains to be
nonrelativistic, and this circumstance leads to many undesirable
consequences. In particular, the particle state, which is defined as a
particle momentum, given at some space-time point, is a concept of
nonrelativistic physics. Concept of the phase space of the particle
positions and momenta is also a nonrelativistic concept. The relativistic
particle state is given in the whole space-time. It is given by the world
line of the particle. Such a concept of the particle state has been
introduced in \cite{R1971}. As a result, ignoring the phase space and using
relativistic concept of the particle state, one succeeded to obtain the
quantum description as a statistical description of stochastic classical
particles. A use of the phase space does not admit one to derive such a
description \cite{M49,F52}.

In this paper we ignore the phase space and use sequentially the
relativistic concept of the particle state. Such an approach admits one to
construct skeleton conception of elementary particles, which realizes a
consistent relativistic dynamics of classical particles in the discrete
space-time geometry.

In non-relativistic physics the state of a physical system is defined as a
set of quantities which are given at a certain moment of time. Equations of
motion determine these quantities at any subsequent moment of time. They
describe the time evolution of the system state. The state and the equations
of motion describing the time evolution of the state are two essential
elements of any non-relativistic physical theory.

As it follows from the definition, a state of a system is given at a certain
time moment. But in relativistic theory a simultaneity is relative. Which
events are synchronous and which are not, it depends on the choice of a
frame of reference. If, for example, one knows a state of a physical system
in a frame of reference $K$, one could describe the state in a frame of
reference $K^{\prime }$ moving relative to $K$ only in the case, when the
equations of motion are known and they can be solved. Thus, in the
relativistic theory the state and the equations of motion are connected
closely. As far as there is no absolute simultaneity in the relativistic
theory, it seems more consistent to define the state of a system not at a
given moment but over all space-time. In this case the concept of state will
include the law of evolution of the physical system. The equations of motion
are treated now as constraints imposed on the admissible states of the
considered system.

Not all possible states are realized. Only those states are realized, which
satisfy certain equations. We shall call them the constraint equations. In
reality they are the same equations of motion but now they do not describe
the time evolution of the state but they are restrictions which choose the
physically allowable states from all virtual ones.

In short, in the non-relativistic theory the unique division of the physical
phenomena description into states and equations of motion corresponds to the
unique division of space-time into space and time. In the relativistic
theory, where the division of space-time into space and time is conventional
and not unique, the division of the physical phenomena description into
states and equations of motion is not unique either. The physical system
state defined over all space-time corresponds much better to the indivisible
space-time.

The manner of division of the physical system description into states and
equations of motion is unimportant for the dynamics of deterministic
particles, but it is important for dynamics of indeterministic particles,
when one uses a statistical description. Any statistics is a calculus of
states. It is important for statistics what is understood under "state". In
general, a statistics that corresponds to a different division of the
description of a physical system into states and equations of motion leads
to different results. The concept of the state density is the main concept
of a statistical description.

In the nonrelativistic physics the state density $\rho $ is defined by the
relation%
\begin{equation}
dN=\rho dV  \label{b1.1}
\end{equation}%
where the state density $\rho $ is the proportionality coefficient between
the infinitesimal 3-volume $dV$ and the number $dN$ of particles in this
3-volume. In the nonrelativistic case the state density $\rho $ is either
3-scalar, or a time component of some timelike 4-vector. Besides, the
nonrelativistic state density $\rho $ is nonnegative. At a proper
normalization the state density $\rho $ may be interpreted as a probability
density of the particle detection at some space point.

In the relativistic physics the state density $j^{k}$ at some space-time
point $x$ is defined as a proportionality coefficient $j^{k}$ between the
infinitesimal 4-vector area $dS_{k}$ and the flux $dF$ of oriented world
lines, crossing this area 
\begin{equation}
dF=j^{k}dS_{k}  \label{b1.2}
\end{equation}%
In the relativistic case the state density is a 4-vector $j^{k}$, $k=0,1,2,3$%
. Its time component $j^{0}$ is positive for a particle and it is negative
for an antiparticle. The time component $j^{0}$ cannot be interpreted as a
probability density, because $j^{0}$ may have any sign. If $j^{0}\left(
x\right) =0$, it does not mean, that the number of particle at the point $x$
is equal to zero. It means only, that the number of particles and the number
of antiparticles at the point $x$ are equal.

In general, a use of nonrelativistic terms at description of relativistic
particles leads to misunderstandings. In the nonrelativistic physics a
pointlike particle $\mathcal{P}$ (a point in 3-space) is considered as a
real physical object, whereas its world line $\mathcal{L}$ is considered as
a property of the pointlike particle $\mathcal{P}$ (its history). In the
relativistic physics the world line $\mathcal{L}$ is a physical object,
whereas the pointlike particle $\mathcal{P}$ is a property of the physical
object $\mathcal{L}$ (intersection of $\mathcal{L}$ with the plane $x^{0}=$
const).

It is useful to introduce a special term for the physical object $\mathcal{L}
$. We use the term "emlon". It is a reading of abbreviation "ML" of Russian
term world line (WL). The point $\mathcal{P}$ of intersection of the emlon
with the surface $x^{0}=$ const is called semlon (or esemlon). It is a
reading of abbreviation SML, which means in Russian "section of world line".
Semlon is a collective concept with respect to concepts of particle and
antiparticle. Particle and antiparticle are two different states of semlon.
In the nonrelativistic physics particle and antiparticle are considered
usually as two different physical objects, and this circumstance is
important in the quantum field theory.

The second quantization of a scalar field is produced usually in terms of
particles and antiparticles, which are considered as independent physical
objects. It leads to nonstationary vacuum state, virtual particles, a use of
a perturbation theory and other exotic results. The second quantization in
terms of emlons (in terms of world lines, considered as physical objects)
leads to a stationary vacuum and to a possibility of a quantization without
recourse to perturbation theory \cite{R1972}.

In general, in the framework of the relativistic physics one uses sometimes
nonrelativistic language, and this prevents from a consistent use of
relativistic physics. For instance, one says: "World line of a particle and
world line of its antiparticle disappear at a collision." From consequent
relativistic viewpoint the same statement should be presented as follows:
"If emlon changes its directivity in the time direction, then one branch of
the emlon describes a particle, whereas another branch describes an
antiparticle." It is different mathematical technique, which is placed
behind the two expressions, describing the same situation.

\label{000}In the nonrelativistic physics the state of a pointlike particle
is given as a point in the phase space. The particle position and the
particle momentum, are given at some time moment, and they describe the
particle state. Besides, one ascribes to the particle its mass and its
charge. Such a description of the pointlike particle state is the same in
the nonrelativistic theory and in the relativistic one.

A use of the phase space supposes, that the particle motion is described by
a smooth continuous world line $x^{k}=x^{k}\left( \tau \right) $, $k=0,1,2,3$
and $\tau $ is a parameter along the world line. It supposes, that the
momentum $p_{k}$ can be defined as limit 
\begin{equation}
p_{k}\left( \tau \right) =\frac{mg_{kl}u^{l}\left( \tau \right) }{\sqrt{%
g_{js}u^{j}\left( \tau \right) u^{s}\left( \tau \right) }},\qquad
u^{l}\left( \tau \right) =\lim_{d\tau \rightarrow 0}\frac{x^{l}\left( \tau
+d\tau \right) -x^{l}\left( \tau \right) }{d\tau }=\frac{du^{l}\left( \tau
\right) }{d\tau }  \label{a1.1}
\end{equation}%
If we consider a particle motion in the microcosm, we cannot be sure, that
the world line is smooth and continuous. Discrete space-time geometry may
violate smoothness of the world line. Furthermore, it is known from
experiments with microparticles (electrons and elementary particles) that
their motion is stochastic, i.e. it cannot be described by a smooth
continuous world line. It is of no importance, whether such a motion is
explained by a discrete space-time geometry or by the quantum nature of
microparticle. In these cases the limit (\ref{a1.1}) does not exist, and one
cannot introduce the phase space, founded on a use of the limit (\ref{a1.1}).

In the case of indeterministic particles we are forced to refuse from a use
of the limit (\ref{a1.1}) and define the pointlike particle state by two
points $P_{0}$, $P_{1}$ in the space-time. The vector $\mathbf{P}_{0}\mathbf{%
P}_{1}$ is an ordered set $\left\{ P_{0},P_{1}\right\} $ of two points $%
P_{0},P_{1}$. The vector $\mathbf{P}_{0}\mathbf{P}_{1}$ describes a
geometrical momentum of the particle, and its length $\mu =\left\vert 
\mathbf{P}_{0}\mathbf{P}_{1}\right\vert $ is its geometrical mass. The usual
4-momentum $\mathbf{p}$ and the usual mass $m$ of the particle are connected
with geometrical quantities by the relations 
\begin{equation}
\mathbf{p}=bc\mathbf{P}_{0}\mathbf{P}_{1},\qquad m=b\mu =b\left\vert \mathbf{%
P}_{0}\mathbf{P}_{1}\right\vert ,\qquad g^{kl}p_{k}p_{l}=m^{2}c^{2}
\label{a1.2}
\end{equation}%
where $b$ is some universal constant, and $c$ is the speed of the light.

Such a generalization of concept of a world line follows from the fact, that
in the proper Euclidean geometry a smooth line is defined as a limit of the
broken straight line, when length of its straight links tends to zero. If
the limit (\ref{a1.1}) does not exist, we are forced to use the broken line
instead of smooth line. In other words, we are forced to use the world chain
instead of a world line.

Experiments show, that the elementary particle motion in the microcosm is
stochastic, and the elementary particles are indeterministic particles. In
this case the limit (\ref{a1.1}) cannot be used, and an indeterministic
pointlike particle is described by the world chain $\mathcal{C}$ (instead of
the world line)%
\begin{equation}
\mathcal{C=}\dbigcup\limits_{s}\mathbf{P}_{s}\mathbf{P}_{s+1}\qquad
\left\vert \mathbf{P}_{k}\mathbf{P}_{k+1}\right\vert =\left\vert \mathbf{P}%
_{k+1}\mathbf{P}_{k+2}\right\vert \quad k=...-1,0,1,...  \label{a1.3}
\end{equation}%
The world chain is described as a set of points $\left\{ P_{s}\right\} $. It
is of no importance, whether there are another points between the points $%
P_{s}$ and $P_{s+1}$, which belong to the chain. In other words, it is
unessential, whether the world chain is used in a continuous geometry or in
a discrete one. Such a definition of the pointlike particle state can be
used in the case of possible discreteness of the space-time geometry.
Besides, this definition does not contain a reference to a coordinate
system. This definition does not need an existence of the limit (\ref{a1.1}%
). World chain may contain links of a finite length. In such a situation one
cannot introduce concept of the phase space in that form, which is used in
the nonrelativistic physics, where the space-time geometry is considered to
be continuous.

If the link length is small enough, from macroscopic viewpoint, it can be
considered as infinitesimal, the world chain can be approximated by a
continuous world line, and the mass can be ascribed to the particle world
line as an external parameter. In this case one can return to the case (\ref%
{a1.1}), and the concept of the phase space can be introduced.

If the particle is free, the adjacent links of the world chain are
equivalent (equal), and the point $P_{s+2}$ is defined via the points $P_{s}$%
, $P_{s+1}$ by means of equations 
\begin{equation}
2\left\vert \mathbf{P}_{s}\mathbf{P}_{s+1}\right\vert =\left\vert \mathbf{P}%
_{s}\mathbf{P}_{s+2}\right\vert ,\quad \left\vert \mathbf{P}_{s}\mathbf{P}%
_{s+1}\right\vert =\left\vert \mathbf{P}_{s+1}\mathbf{P}_{s+2}\right\vert
,\qquad s=0,\pm 1,\pm 2,...  \label{a1.4}
\end{equation}%
In the space-time of Minkowski the equations (\ref{a1.4}) has a unique
solution for the point $P_{s+2}$, provided all links are timelike.

In the discrete (physical) space-time geometry, where geometry is defined
completely by the world function $\sigma \left( P_{0},P_{1}\right) =\frac{1}{%
2}\left\vert \mathbf{P}_{0}\mathbf{P}_{1}\right\vert ^{2}$, the equations (%
\ref{a1.4}) has many solutions, in general, even for timelike links. In this
case the phase space cannot be introduced, because the limit (\ref{a1.1})
does not exist. In other words, one cannot use nonrelativistic concept of
the particle state in the microcosm.

The rule (\ref{a1.4}) of the world chain construction for a free particle
coincides with the rule of straight line construction by means of a compass
in the proper Euclidean geometry. In the discrete space-time geometry the
rule (\ref{a1.4}) is formulated as follows. The adjacent links are
equivalent $\mathbf{P}_{s}\mathbf{P}_{s+1}$eqv$\mathbf{P}_{s+1}\mathbf{P}%
_{s+2}$.

In the proper Euclidean geometry $\mathcal{G}_{\mathrm{E}}$ the equivalence
of two vectors $\mathbf{P}_{0}\mathbf{P}_{1}$ and $\mathbf{Q}_{0}\mathbf{Q}%
_{1}$ is defined as follows. Vectors $\mathbf{P}_{0}\mathbf{P}_{1}$ and $%
\mathbf{Q}_{0}\mathbf{Q}_{1}$ are equivalent ($\mathbf{P}_{0}\mathbf{P}_{1}$%
eqv $\mathbf{Q}_{0}\mathbf{Q}_{1}$), if vectors $\mathbf{P}_{0}\mathbf{P}%
_{1} $ and $\mathbf{Q}_{0}\mathbf{Q}_{1}$ are in parallel $\left( \mathbf{P}%
_{0}\mathbf{P}_{1}\uparrow \uparrow \mathbf{Q}_{0}\mathbf{Q}_{1}\right) $
and their lengths $\left\vert \mathbf{P}_{0}\mathbf{P}_{1}\right\vert $ and $%
\left\vert \mathbf{Q}_{0}\mathbf{Q}_{1}\right\vert $ are equal.
Mathematically the two conditions are written in the form%
\begin{equation}
\left( \mathbf{P}_{0}\mathbf{P}_{1}\uparrow \uparrow \mathbf{Q}_{0}\mathbf{Q}%
_{1}\right) :\qquad \left( \mathbf{P}_{0}\mathbf{P}_{1}.\mathbf{Q}_{0}%
\mathbf{Q}_{1}\right) =\left\vert \mathbf{P}_{0}\mathbf{P}_{1}\right\vert
\cdot \left\vert \mathbf{Q}_{0}\mathbf{Q}_{1}\right\vert  \label{a1.5}
\end{equation}

\begin{equation}
\left\vert \mathbf{P}_{0}\mathbf{P}_{1}\right\vert =\left\vert \mathbf{Q}_{0}%
\mathbf{Q}_{1}\right\vert ,\qquad \left\vert \mathbf{P}_{0}\mathbf{P}%
_{1}\right\vert =\sqrt{2\sigma \left( P_{0},P_{1}\right) }  \label{a1.6}
\end{equation}%
where $\left( \mathbf{P}_{0}\mathbf{P}_{1}.\mathbf{Q}_{0}\mathbf{Q}%
_{1}\right) $ is the scalar product of two vectors. It is defined by the
relation%
\begin{equation}
\left( \mathbf{P}_{0}\mathbf{P}_{1}.\mathbf{Q}_{0}\mathbf{Q}_{1}\right)
=\sigma \left( P_{0},Q_{1}\right) +\sigma \left( P_{1},Q_{0}\right) -\sigma
\left( P_{0},Q_{0}\right) -\sigma \left( P_{1},Q_{1}\right)  \label{a1.7}
\end{equation}%
Here $\sigma $ is the world function of the proper Euclidean geometry $%
\mathcal{G}_{\mathrm{E}}$. The length $\left\vert \mathbf{PQ}\right\vert $
of vector $\mathbf{PQ}$ is defined by the relation%
\begin{equation}
\left\vert \mathbf{PQ}\right\vert =\rho \left( P,Q\right) =\sqrt{2\sigma
\left( P,Q\right) }  \label{a1.8}
\end{equation}%
The definition (\ref{a1.7}) contains neither coordinate system, nor
dimension of the space, and it may be used in any physical geometry, i.e. in
the geometry described by the world function completely.

Using relations (\ref{a1.5}) - (\ref{a1.8}), one can write the equivalence
condition in the form%
\begin{eqnarray}
\mathbf{P}_{0}\mathbf{P}_{1}\mathrm{eqv}\mathbf{Q}_{0}\mathbf{Q}_{1}
&:&\quad \sigma \left( P_{0},Q_{1}\right) +\sigma \left( P_{1},Q_{0}\right)
-\sigma \left( P_{0},Q_{0}\right) -\sigma \left( P_{1},Q_{1}\right) =2\sigma
\left( P_{0},P_{1}\right)  \notag \\
\wedge \sigma \left( P_{0},P_{1}\right) &=&\sigma \left( Q_{0},Q_{1}\right)
\label{a1.9}
\end{eqnarray}%
If $P_{0}=P_{s},\ Q_{0}=P_{1}=P_{s+1}$ and $Q_{1}=P_{s+2}$, the relations (%
\ref{a1.9}) take the form%
\begin{equation}
\mathbf{P}_{s}\mathbf{P}_{s+1}\mathrm{eqv}\mathbf{P}_{s+1}\mathbf{P}%
_{s+2}:\quad \sigma \left( P_{s},P_{s+2}\right) =4\sigma \left(
P_{s},P_{s+1}\right) \wedge \sigma \left( P_{s},P_{s+1}\right) =\sigma
\left( P_{s+1},P_{s+2}\right)  \label{a1.10}
\end{equation}%
which coincides with (\ref{a1.4})

The equivalence relation (\ref{a1.9}) is used in any physical geometry.

\section{Statistical description of indeterministic \newline
particles}

In the beginning of the twentieth century it was natural to think, that the
quantum particles are simply indeterministic (stochastic) particles,
something like Brownian particles. There were attempts to obtain quantum
mechanics as a statistical description of stochastically moving particles 
\cite{M49,F52}. However, these attempts failed because a \textit{%
probabilistic conception of the statistical description} was used.

Statistical description is used in physics for description of
indeterministic particles (or systems), when there are no dynamic equations,
or initial conditions are indefinite. One considers statistical ensemble of
indeterministic particles, i.e. many independent similar particles. It
appears, that there are dynamic equations for the statistical ensemble $%
\mathcal{E}$ of indeterministic particles, although there are no dynamic
equations for a single indeterministic particle, which is a constituent of
this statistical ensemble $\mathcal{E}$. Consideration of the statistical
ensemble as a dynamic system is a \textit{dynamic conception of the
statistical description }(DCSD). It is a primordial conception of
statistical description. A use of DCSD is founded on independence of
constituents of the statistical ensemble. Random components of motion are
compensated due to their independence, whereas regular components of motion
are accumulated.

In the nonrelativistic physics \textit{the probabilistic conception of the
statistical description} (PCSD) is used. PCSD is used successfully, for
instance, for description of Brownian motion. In PCSD one traces the motion
of the point in the phase space. The point represents the state of
indeterministic particle, and a motion the point in the phase space is
described by the probability transition. Attempts of obtaining the quantum
mechanics as a result of description in the framework PCSD failed, because
PCSD is a nonrelativistic description, whereas the nonrelativistic quantum
mechanics is a relativistic construction, and the quantum mechanics should
be obtained as a statistical description in terms of DCSD.

But why is the nonrelativistic quantum mechanics a relativistic
construction? Because the stochastic component of the quantum particle
motion may be relativistic, and one has to use the dynamic conception of
statistical description (DCSD), which does not use the nonrelativistic
concept of the phase space.

Indeed, in terms of DCSD one succeeded to obtain the quantum mechanics as a
statistical description of stochastically moving particles \cite%
{R1971,R73a,R73,R2006}. The action for the statistical ensemble $\mathcal{E}%
\left[ \mathcal{S}_{\mathrm{st}}\right] $ of free indeterministic particles $%
\mathcal{S}_{\mathrm{st}}$ is written in the form%
\begin{equation}
\mathcal{A}_{\mathcal{E}\left[ \mathcal{S}_{\mathrm{st}}\right] }\left[ 
\mathbf{x},\mathbf{u}\right] =\int \dint\limits_{V_{\xi }}\left\{ \frac{m}{2}%
\mathbf{\dot{x}}^{2}+\frac{m}{2}\mathbf{u}^{2}-\frac{\hbar }{2}\mathbf{%
\nabla u}\right\} dtd\mathbf{\xi },\qquad \mathbf{\dot{x}\equiv }\frac{d%
\mathbf{x}}{dt}  \label{d1.5}
\end{equation}%
Independent variables $\mathbf{\xi }=\left\{ \xi _{1},\xi _{2},\xi
_{3}\right\} $ label constituents $\mathcal{S}_{\mathrm{st}}$ of the
statistical ensemble. The dependent variable $\mathbf{x}=\mathbf{x}\left( t,%
\mathbf{\xi }\right) $ describes the regular component of the particle
motion. The variable $\mathbf{u}=\mathbf{u}\left( t,\mathbf{x}\right) $
describes the mean value of the stochastic velocity component, $\hbar $ is
the quantum constant. The second term in (\ref{d1.5}) describes the kinetic
energy of the stochastic velocity component. The third term describes
interaction between the stochastic component $\mathbf{u}\left( t,\mathbf{x}%
\right) $ and the regular component $\mathbf{\dot{x}}\left( t,\mathbf{\xi }%
\right) $. The operator 
\begin{equation}
\mathbf{\nabla =}\left\{ \frac{\partial }{\partial x^{1}},\frac{\partial }{%
\partial x^{2}},\frac{\partial }{\partial x^{3}}\right\}  \label{d1.5a}
\end{equation}%
is defined in the space of coordinates $\mathbf{x}$. Dynamic equations for
the dynamic system $\mathcal{E}\left[ \mathcal{S}_{\mathrm{st}}\right] $ are
obtained as a result of variation of the action (\ref{d1.5}) with respect to
dynamic variables $\mathbf{x}$ and $\mathbf{u}$.

The action for a single indeterministic particle $\mathcal{S}_{\mathrm{st}}$
has the form%
\begin{equation}
\mathcal{A}_{\mathcal{S}_{\mathrm{st}}}\left[ \mathbf{x},\mathbf{u}\right]
=\int \dint\limits_{V_{\xi }}\left\{ \frac{m}{2}\mathbf{\dot{x}}^{2}+\frac{m%
}{2}\mathbf{u}^{2}-\frac{\hbar }{2}\mathbf{\nabla u}\right\} dt,\qquad 
\mathbf{\dot{x}\equiv }\frac{d\mathbf{x}}{dt}  \label{a1.11}
\end{equation}%
This action is not correctly defined, because operator $\mathbf{\nabla }$ is
defined on 3D-space of coordinates $\mathbf{x}=\left\{
x^{1},x^{2},x^{3}\right\} $, whereas in the action functional (\ref{a1.11})
the variable $\mathbf{x}$ is used only on one-dimensional set. It means that
there are no dynamic equations for the particle $\mathcal{S}_{\mathrm{st}}$,
and the particle $\mathcal{S}_{\mathrm{st}}$ is a stochastic (not dynamic)
system. However, the action functional (\ref{d1.5}) is well defined, and
dynamic equations exist for the statistical ensemble $\mathcal{E}\left[ 
\mathcal{S}_{\mathrm{st}}\right] $, although dynamic equations do not exist
for constituents of this statistical ensemble.

Variation of the action (\ref{d1.5}) leads to dynamic equations%
\begin{equation}
\delta \mathbf{u:\qquad }m\rho \mathbf{u}+\frac{\hbar }{2}\mathbf{\nabla }%
\rho =0,\qquad \mathbf{u}=-\frac{\hbar }{2m}\mathbf{\nabla }\ln \rho
\label{d1.7}
\end{equation}%
\begin{equation}
\delta \mathbf{x:\qquad }m\frac{d^{2}\mathbf{x}}{dt^{2}}=\mathbf{\nabla }%
\left( \frac{m}{2}\mathbf{u}^{2}-\frac{\hbar }{2}\mathbf{\nabla u}\right)
\label{d1.8}
\end{equation}%
where%
\begin{equation}
\rho =\frac{\partial \left( \xi _{1},\xi _{2},\xi _{3}\right) }{\partial
\left( x^{1},x^{2},x^{3}\right) }=\left( \frac{\partial \left(
x^{1},x^{2},x^{3}\right) }{\partial \left( \xi _{1},\xi _{2},\xi _{3}\right) 
}\right) ^{-1}  \label{d1.7a}
\end{equation}

After proper change of variables the dynamic equations are reduced to the
equation \cite{R2006}%
\begin{equation}
i\hbar \partial _{0}\psi +\frac{\hbar ^{2}}{2m}\mathbf{\nabla }^{2}\psi +%
\frac{\hbar ^{2}}{8m}\mathbf{\nabla }^{2}s_{\alpha }\cdot \left( s_{\alpha
}-2\sigma _{\alpha }\right) \psi -\frac{\hbar ^{2}}{4m}\frac{\mathbf{\nabla }%
\rho }{\rho }\mathbf{\nabla }s_{\alpha }\sigma _{\alpha }\psi =0
\label{d6.14}
\end{equation}%
where $\psi $ is the two component complex wave function%
\begin{equation}
\rho =\psi ^{\ast }\psi ,\qquad s_{\alpha }=\frac{\psi ^{\ast }\sigma
_{\alpha }\psi }{\rho },\qquad \alpha =1,2,3  \label{d6.7}
\end{equation}%
$\sigma _{\alpha }$ are $2\times 2$ Pauli matrices%
\begin{equation}
\sigma _{1}=\left( 
\begin{array}{cc}
0 & 1 \\ 
1 & 0%
\end{array}%
\right) ,\qquad \sigma _{2}=\left( 
\begin{array}{cc}
0 & -i \\ 
i & 0%
\end{array}%
\right) ,\qquad \sigma _{3}=\left( 
\begin{array}{cc}
1 & 0 \\ 
0 & -1%
\end{array}%
\right) ,  \label{d6.8}
\end{equation}

If components $\psi _{1}$ and $\psi _{2}$ are linear dependent $\psi =\left( 
\begin{array}{c}
a\psi _{1} \\ 
b\psi _{1}%
\end{array}%
\right) $, $a,b=\mathrm{const}$, then $\mathbf{\ s}=\mathrm{const}$. Two
last terms of the equation (\ref{d6.14}) vanish, and the equation turns to
the Schr\"{o}dinger equation%
\begin{equation}
i\hbar \partial _{0}\psi +\frac{\hbar ^{2}}{2m}\mathbf{\nabla }^{2}\psi =0
\label{a1.12}
\end{equation}

Thus, the Schr\"{o}dinger equation and interpretation of the quantum
mechanics appear from the dynamical system $\mathcal{E}\left[ \mathcal{S}_{%
\mathrm{st}}\right] $, described by the action functional (\ref{d1.5}). This
fact seems rather unexpected, because in quantum mechanics the wave function
is considered as a specific quantum object, which has no analog in classical
physics. In reality, the wave function is simply a way of description of
ideal continuous medium \cite{R1999}. You may describe an ideal fluid in
terms of hydrodynamic variables: density $\rho $ and velocity $\mathbf{v}$%
\textbf{. }You may describe an ideal fluid in terms of the wave function.
The statistical ensemble $\mathcal{E}\left[ \mathcal{S}_{\mathrm{st}}\right] 
$ is a dynamic system of the type of continuous medium. The two
representations of dynamic equations for the dynamic system $\mathcal{E}%
\left[ \mathcal{S}_{\mathrm{st}}\right] $ can be transformed one into
another.

It is well known, that the Schr\"{o}dinger equation can be written in the
hydrodynamic form of Madelung-Bohm \cite{M26,B52}. The wave function $\psi $
is presented in the form 
\begin{equation}
\psi =\sqrt{\rho }\exp \left( i\varphi /\hbar \right)  \label{a2.1}
\end{equation}%
Substituting (\ref{a2.1}) in the Schr\"{o}dinger equation (\ref{a1.12}), one
obtains two real equations for dynamical variables $\rho $ and $\varphi $.
Taking gradient from the equation for $\varphi $ and introducing designation 
\begin{equation}
\mathbf{v=-}\frac{\hbar }{m}\mathbf{\nabla }\varphi ,\qquad \func{curl}%
\mathbf{v}=0  \label{a2.2}
\end{equation}%
one obtains four equations of the hydrodynamical type 
\begin{equation}
\frac{\partial \rho }{\partial t}+\mathbf{\nabla }\left( \rho \mathbf{v}%
\right) =0,\qquad \frac{d\mathbf{v}}{dt}\equiv \frac{\partial \mathbf{v}}{%
\partial t}+\left( \mathbf{v\nabla }\right) \mathbf{v}=-\frac{1}{m}\mathbf{%
\nabla }U_{\mathrm{B}}  \label{d1.14}
\end{equation}%
where $U_{\mathrm{B}}$ is the Bohm potential, defined by the relation 
\begin{equation}
U_{\mathrm{B}}=U\left( \rho ,\mathbf{\nabla }\rho ,\mathbf{\nabla }^{2}\rho
\right) =\frac{\hbar ^{2}}{8m\rho }\left( \frac{\left( \mathbf{\nabla }\rho
\right) ^{2}}{\rho }-2\mathbf{\nabla }^{2}\rho \right) =\mathbf{-}\frac{%
\hbar ^{2}}{2m\sqrt{\rho }}\mathbf{\nabla }^{2}\sqrt{\rho }  \label{d4.3}
\end{equation}%
Hydrodynamic equations (\ref{d1.14}) can be easily obtained from equations (%
\ref{d1.7}), (\ref{d1.8}). To obtain representation of equations (\ref{d1.14}%
), (\ref{d4.3}) in terms of wave function, one needs to integrate these
equations, because they have been obtained by means of differentiation of
the Schr\"{o}dinger equation. This integration can be easily produced, if
the condition (\ref{a2.2}) takes place and the fluid flow is non-rotational.

In the general case of vortical flow the integration is more complicated.
Nevertheless this integration has been produced \cite{R1999}, and one
obtains a more complicated equation (\ref{d6.14}), where two last terms
describe vorticity of the flow. The Schr\"{o}dinger equation (\ref{a1.12})
is a special case of the more general equation (\ref{d6.14}).

Note that the equation (\ref{d6.14}) is not linear, although it is invariant
with respect to transformation 
\begin{equation}
\psi \rightarrow \tilde{\psi}=A\psi ,\qquad A=\text{const}  \label{a2.3}
\end{equation}%
which admits one to normalize the wave function to any nonnegative quantity.
This property describes independence of the statistical ensemble on the
number of its constituents

\label{0beg1}Description of the pair production is obtained in the
relativistic version of the action functional (\ref{d1.5}). This action has
the form%
\begin{equation}
\mathcal{A}_{\mathcal{E}\left[ \mathcal{S}_{\mathrm{st}}\right] }\left[ x%
\mathbf{,}\kappa \right] =-\int \dint\limits_{V_{\mathbf{\xi }}}mcK\sqrt{%
g_{ik}\dot{x}^{i}\dot{x}^{k}}\rho _{0}\left( \mathbf{\xi }\right) d\tau d%
\mathbf{\xi },\qquad \mathbf{\dot{x}\equiv }\frac{d\mathbf{x}}{d\tau }
\label{d2.8}
\end{equation}%
\begin{equation}
K=\sqrt{1+\lambda ^{2}\left( g_{kl}\kappa ^{k}\kappa ^{l}+\partial
_{k}\kappa ^{k}\right) },\qquad \lambda =\frac{\hbar }{mc}  \label{d2.9}
\end{equation}%
where $x=\left\{ x^{k}\right\} =\left\{ x^{k}\left( \tau ,\mathbf{\xi }%
\right) \right\} ,$ $k=0,1,2,3$ are dependent variables. The quantity $%
g_{kl}=$diag$\left\{ c^{2},-1,-1,-1\right\} $ is the metric tensor. The
independent variables $\mathbf{\xi }=\left\{ \xi _{1},\xi _{2},\xi
_{3}\right\} $ label the particles of the statistical ensemble. The
dependent variables $\kappa ^{k}=\kappa ^{k}\left( x\right) $, $k=0,1,2,3$
form some force field, connected with the stochastic component of the
particle 4-velocity, and $\lambda $ is the Compton wave length of the
particle. Connection of the field $\kappa ^{k}$ with the mean value $%
u^{k}\left( t,\mathbf{x}\right) =u^{k}\left( x\right) $ of stochastic
component of 4-velocity has the form%
\begin{equation}
\kappa ^{k}=\frac{m}{\hbar }u^{k},\qquad k=0,1,2,3  \label{d2.7}
\end{equation}

In the nonrelativistic approximation one may neglect the temporal component $%
\kappa ^{0}=\frac{m}{\hbar }u^{0}$ with respect to the spatial one $\mathbf{%
\kappa }=\frac{m}{\hbar }\mathbf{u}.$ Setting $\tau =t$ $=x^{0}$ in (\ref%
{d2.8}), (\ref{d2.9}), we obtain the action (\ref{d1.5}) instead of (\ref%
{d2.8}).

To describe the pair production, the world line is to have a possibility of
turn in the time direction. At the turning point the world line has to be
spacelike and the radical in (\ref{d2.8}) must be imaginary. It is possible,
only if the quantity (\ref{d2.9}) is imaginary also. It means, the effective
mass $mK$ is to be imaginary. The quantity $K$ may be imaginary, if the
field $\kappa ^{k}$ have proper values. It means that the stochastic
component of the particle motion is responsible for the pair production
(turn of the world line in the time direction). \label{0end1}

Representation of quantum mechanics as a statistical description of
classical indeterministic particles admits one to interpret all quantum
relations in terms of statistical description. This interpretation
distinguishes in some clauses from the conventional interpretation of
quantum mechanics.

In any statistical description there are two different kinds of measurement,
which have different properties. Massive measurement (M-measurement) is
produced over all constituents of the statistical ensemble. A result of
M-measurement of the quantity $R$ is a distribution of the quantity $R$,
which can be predicted as a result of solution of dynamic equations for the
statistical ensemble.

Single measurement (S-measurement) is produced over one of constituents of
the statistical ensemble. A result of S-measurement of the quantity $R$ is
some random value of the quantity $R$, which cannot be predicted by the
theory. In the Copenhagen interpretation of the quantum mechanics the wave
function is supposed to describe a single particle (but not a statistical
ensemble of particles). As a result there is only one type of measurement,
which is considered sometimes as a M-measurement and sometimes as a
S-measurement. As far as M-measurement and S-measurement have different
properties, such an identification is a source of numerous contradictions
and paradoxes \cite{R2006a}.

Representation of quantum mechanics as a statistical description of the
indeterministic particles motion has two important consequences: (1)
elimination of quantum principles as laws of nature, (2) problem of
primordial stochastic motion of free particles.

\section{Multivariant space-time geometry as a \newline
corollary of existence of indeterministic \newline
particles.}

Reduction of number of physical principles means an increase of the quality
of the physical theory. Explanation of quantum effects by means of a
stochasticity of free particle motion sets the question of the nature of
this stochasticity. The \textit{motion of a free particle is determined by
properties of the space-time geometry}. The free particle motion is
deterministic in the space-time of Minkowski. An indeterministic motion of
free particles is possible only in multivariant space-time geometries. Such
geometries were unknown in the twentieth century, and explanation of quantum
effects by a stochasticity of particle motion seemed to be impossible.

The multivariant (physical) geometry is nonaxiomatizable, in general. It
means, that statements of the multivariant geometry cannot be deduced from
axiomatics. In the twentieth century only axiomatizable geometries were
known. Mathematicians, who were responsible for investigation and creation
of geometries, believed that any geometry is to be a logical construction.
Hence, any geometry is to be axiomatizable.

In general, there were mathematicians \cite{M28,B53}, who believed that the
geometry may be a distance geometry, which is described by the distance
function between any two points of the space. However, it was not known, how
to construct geometrical objects in the distance geometry. The distance
geometry appeared to be ineffective, and at description of the space-time
the mathematicians ignored the distance geometry, as well as the metric
geometry, which a special case of the distance geometry.

Situation was changed cardinally, when a way of the geometrical objects
construction has been invented. It is the deformation principle \cite{R2007}%
. One takes a geometrical object of the proper Euclidean geometry and
describes it in terms of the Euclidean distance function $\rho _{\mathrm{E}}$
(or in terms of the Euclidean world function $\sigma _{\mathrm{E}}=\frac{1}{2%
}\rho _{\mathrm{E}}^{2}$). Substituting Euclidean distance function $\rho _{%
\mathrm{E}}$ by the distance function $\rho $ of the geometry in question $%
\mathcal{G}$, one obtains the geometrical object of the geometry $\mathcal{G}
$. Although the deformation principle has been published in explicit form
only in 2007, in fact it was used ab origine of the physical geometry
construction \cite{R2002}.

As far as the formal logic is not used at the construction of geometric
objects of the physical geometry, the obtained physical geometry is
multivariant and nonaxiomatizable, in general. It means, that solving
equations (\ref{a1.9}) at given vector $\mathbf{P}_{0}\mathbf{P}_{1}$ and
given point $Q_{0}$, one obtains, in general, many solutions for the vector $%
\mathbf{Q}_{0}\mathbf{Q}_{1}$. It is possible also such a situation, when
equations (\ref{a1.9}) have no solution.

Multivariant space-time geometry made impossible a use of the limit (\ref%
{a1.1}) for a construction of the phase space, and nonrelativistic concept
of the particle state becomes impossible for description of elementary
particles. On the other hand, being a reason of the free particles motion
stochasticity, the multivariant space-time geometry becomes to be
interpreted as a reason of quantum effects \cite{R91}. Let us stress, that
the obtaining of the Schr\"{o}dinger equation as corollary of the
multivariant space-time geometry appeared to be possible only at a use of
the relativistic concept of the pointlike particle state (\ref{a1.2}), (\ref%
{a1.3}). Only in this case the free particle motion depends on the particle
mass. Indeed, describing a free particle motion, the Schr\"{o}dinger
equation contains the particle mass, whereas the classic deterministic
motion of a free particle is the same for particles of any mass. The length
of links (\ref{a1.2}) of the world chain (\ref{a1.3}) is essential for the
stochastic component of the particle motion.

\section{Skeleton conception of elementary particles}

After the paper \cite{R91} the role of the space-time geometry increased in
the theory of elementary particles, because in fact the quantum principles
were replaced by the multivariant space-time geometry. It became clear, that
constructing a theory of elementary particles, one should use relativistic
concept of the particle state.

In the case, when the particle is not pointlike, its state is described by
its skeleton $\mathcal{P}_{n}=\left\{ P_{0},P_{1},...,P_{n}\right\} $, which
is a set of $(n+1)$ space-time points. $n>1$ is some integer number. These
points are connected rigidly. In the case of a pointlike particle the
skeleton consists of two points. The skeleton $\mathcal{P}_{n}$ is a natural
generalization of the skeleton of the pointlike particle on the case of a
composite particle. Motion of any particle is described by the world chain,
consisting of connected skeletons \cite{R2008}. ...$\mathcal{P}_{n}^{\left(
0\right) },\mathcal{P}_{n}^{\left( 1\right) },...,\mathcal{P}_{n}^{\left(
s\right) }...$

\begin{equation}
\mathcal{P}_{n}^{\left( s\right) }=\left\{ P_{0}^{\left( s\right)
},P_{1}^{\left( s\right) },..P_{n}^{\left( s\right) }\right\} ,\qquad
s=...0,1,...  \label{b4.8}
\end{equation}%
The adjacent skeletons $\mathcal{P}_{n}^{\left( s\right) },\mathcal{P}%
_{n}^{\left( s+1\right) }$ of the chain are connected by the relations $%
P_{1}^{\left( s\right) }=P_{0}^{\left( s+1\right) }$, $s=...0,1,...$ The
vector $\mathbf{P}_{0}^{\left( s\right) }\mathbf{P}_{1}^{\left( s\right) }=%
\mathbf{P}_{0}^{\left( s\right) }\mathbf{P}_{0}^{\left( s+1\right) }$ is the
leading vector, which determines the world chain direction.

Dynamics of a free elementary particle is determined by the relations%
\begin{equation}
\mathcal{P}_{n}^{\left( s\right) }\mathrm{eqv}\mathcal{P}_{n}^{\left(
s+1\right) }:\qquad \mathbf{P}_{i}^{\left( s\right) }\mathbf{P}_{k}^{\left(
s\right) }\mathrm{eqv}\mathbf{P}_{i}^{\left( s+1\right) }\mathbf{P}%
_{k}^{\left( s+1\right) },\qquad i,k=0,1,..n;\qquad s=...0,1,...
\label{a4.8}
\end{equation}%
which describe equivalence of adjacent skeletons. Equivalence of vectors is
defined by the relations (\ref{a1.9}).

Thus, dynamics of a free elementary particle is described by a system of
algebraic equations (\ref{a4.8}). Specific of dynamics depends on the
elementary particle structure (disposition of particles inside the skeleton)
and on the space-time geometry. Lengths $\left\vert \mathbf{P}_{i}^{\left(
s\right) }\mathbf{P}_{k}^{\left( s\right) }\right\vert $ of vectors $\mathbf{%
P}_{i}^{\left( s\right) }\mathbf{P}_{k}^{\left( s\right) }$ are constant
along the whole world chain. These $n\left( n+1\right) /2$ quantities may be
considered as characteristics of the particle. In the case of pointlike
particle the length $\left\vert \mathbf{P}_{s}\mathbf{P}_{s+1}\right\vert $
of the link $\mathbf{P}_{s}\mathbf{P}_{s+1}$ is the geometrical mass of the
particle. In the case of a more complicated skeletons the meaning of
parameters $\left\vert \mathbf{P}_{i}^{\left( s\right) }\mathbf{P}%
_{k}^{\left( s\right) }\right\vert $ should be investigated.

\textit{Remark. }We are forced to reject from definition of particle state
as some quantities given at some time moment, because some vectors of a
skeleton are timelike, and one cannot find such a coordinate system, where
all points of the skeleton have the same time coordinate. One cannot define
the particle state as points of intersection of several world lines with the
surface $x^{0}=$ const, because the space-time geometry may be discrete and
continuous world lines do not exist.

The system of dynamic equations (\ref{a4.8}) consists of $n\left( n+1\right) 
$ algebraic equations for $nD$ dynamic variables, where $D$ is the dimension
the space-time (the number of coordinates, which are necessary for labelling
of all points of the space-time). If $n\leq D$, the number of dynamic
variables is less, than the number of dynamic equations. In this case we
have a discrimination mechanism, which forbids some skeletons. This
mechanism admits one to explain discrete parameters of elementary particles.
If $n>D+1$, the number of dynamic equations is more than the number of
dynamic variables. In this case there may exist many solutions, and the
particle motion becomes multivariant. Both cases may take place in the
theory of elementary particles.

Dynamic equations (\ref{a4.8}) are written in the coordinateless form, and
this fact is a worth of the dynamic equations (\ref{a4.8}), as far as it
saves from a necessity to consider the coordinate transformations. Dynamic
equations (\ref{a4.8}) are algebraic equations (not differential), and this
fact is also a worth of the theory, because the algebraic equations may be
used even in a discrete space-time geometry.

The first (nontrivial) attempt of a use of the relativistic concept of the
particle state was made. One considered the structure of the Dirac particle
(fermion) \cite{R2008a}. It appeared that the skeleton of the Dirac particle
consists of $n$ points ($n\geq 3$). Its world chain is a spacelike helix
with a timelike axis. Spacelike world lines are impossible in the space-time
geometry of Minkowski.

The Dirac particle is considered in the space-time geometry described by the
world function $\sigma _{\mathrm{d}}$%
\begin{equation}
\sigma _{\mathrm{d}}=\sigma _{\mathrm{M}}+\lambda _{0}^{2}\left\{ 
\begin{array}{lll}
\mathrm{sgn}\left( \sigma _{\mathrm{M}}\right) & \text{if} & \left\vert
\sigma _{\mathrm{M}}\right\vert >\sigma _{0}>0 \\ 
f\left( \sigma _{\mathrm{M}}\right) & \text{if} & \left\vert \text{\ }\sigma
_{\mathrm{M}}\right\vert <\sigma _{0}%
\end{array}%
\right. \qquad \lambda _{0}^{2}=\frac{\hbar }{2bc}  \label{a4.14}
\end{equation}%
where $\sigma _{\mathrm{M}}$ is the world function of the space-time of
Minkowski, $b$ is some universal constant and $\sigma _{0}$ is some
constant. The function $f$ is a monotone nondecreasing function, having
properties $f\left( -\sigma _{0}\right) =-1$, $f\left( \sigma _{0}\right) =1$%
.

The space-time geometry, described by the world function (\ref{a4.14}) is
uniform and isotropic. The part of the world function corresponding to $%
\left\vert \sigma _{\mathrm{M}}\right\vert >\sigma _{0}$ is responsible for
quantum effects of a pointlike particle (Schr\"{o}dinger equation \cite{R91}%
). The part of the world function (\ref{a4.14}), corresponding to $%
\left\vert \sigma _{\mathrm{M}}\right\vert <\sigma _{0}$ is responsible the
structure of a particle with the skeleton consisting of more, than two
points. If $\left\vert f\left( \sigma _{\mathrm{M}}\right) \right\vert
<\left\vert \sigma _{\mathrm{M}}/\sigma _{0}\right\vert $, the spacelike
world chain may have a shape of a helix with a timelike axis.

The case, when 
\begin{equation}
f\left( \sigma _{\mathrm{M}}\right) =\left( \frac{\sigma _{\mathrm{M}}}{%
\sigma _{0}}\right) ^{3}  \label{a2.5}
\end{equation}%
has been investigated. Such a choice of the world function does not pretend
to description of the real space-time. It is only some model, which
correctly describes quantum effects connected with pointlike particles and
tries to investigate, whether spacelike world chain may have a shape of a
helix with a timelike axis. According to semiclassical approximation of the
Dirac equation \cite{R1995,R2004,R2004b} the world line of a \textit{free
classical} Dirac particle has the shape of a helix. Such a shape of the
world line explains existence of a spin. It was interesting, whether the
spin of the Dirac particle can be obtained in the skeleton conception of
elementary particles.

Consideration in \cite{R2008a} confirmed the supposition on the helix world
chain of the Dirac particle (fermion). The skeleton of a fermion is to
contain more, than two points. Besides, some restrictions on disposition of
the skeleton points were obtained. It means that in the skeleton conception
there is a discrimination mechanism responsible for discrete values of
parameters of the elementary particles. Such a discrimination mechanism is
absent in the conventional approach, based on a use of quantum principles.
The obtained results are preliminary, because the simple restriction (\ref%
{a2.5}) on the world function has been used. Nevertheless these results
show, that the skeleton conception admits one to investigate the structure
of elementary particles. The conventional approach, based on quantum
principles, admits one only to ascribe to elementary particles such
phenomenological properties as spin, color, flavour and other, without
explanation how these properties relate to the elementary particle
structure. The quantum approach admits one only to classify elementary
particles by their phenomenological properties and to predict reaction
between the elementary particles on the basis of this classification.

Such a situation reminds situation with investigation of chemical elements.
Periodic system of chemical elements is a phenomenological construction. It
is an attribute of chemistry. Arrangement of atoms of chemical elements is
investigated by physics (quantum mechanics). The periodic system of chemical
elements had been discovered earlier, than researchers began to investigate
atomic structures. However, the periodic system did not help us to create
quantum mechanics and to investigate the atomic structure. The periodic
system and the quantum mechanics are attributes of different sciences. In
the same track the skeleton conception of elementary particles and the
conventional phenomenological approach based on quantum theory are
essentially attributes of different sciences, investigating different sides
of the elementary particles.

\section{About conservation laws}

Conservation laws of energy-momentum and angular momentum take place only in
a uniform and isotropic space-time geometry. The real space-time geometry $%
\mathcal{G}_{\mathrm{r}}$, where all elementary particles move freely, is
not uniform and isotropic, in general. To obtain the conservation laws, we
may consider some fictitious space-time geometry $\mathcal{G}_{\mathrm{f}}$,
which is uniform and isotropic. For instance, we may suppose, that the
space-time geometry $\mathcal{G}_{\mathrm{f}}$ is the geometry of Minkowski $%
\mathcal{G}_{\mathrm{M}}$. Let us describe the particle motion in the
space-time geometry of Minkowski. The particle motion, which is free in the
geometry $\mathcal{G}_{\mathrm{r}}$ ceases to be free in the fictitious
geometry $\mathcal{G}_{\mathrm{M}}$. Some force fields appear. These force
fields appear as a result of mismatch $d$ between the world functions $%
\sigma _{\mathrm{r}}$ and $\sigma _{\mathrm{M}}$%
\begin{equation}
d\left( P,Q\right) =\sigma _{\mathrm{r}}\left( P,Q\right) -\sigma _{\mathrm{M%
}}\left( P,Q\right)  \label{a5.1}
\end{equation}%
Rewriting dynamic equations (\ref{a4.8}) in terms of the world function $%
\sigma _{\mathrm{M}}$, one obtains%
\begin{equation}
\left( \mathbf{P}_{i}^{\left( s\right) }\mathbf{P}_{k}^{\left( s\right) }.%
\mathbf{P}_{i}^{\left( s+1\right) }\mathbf{P}_{k}^{\left( s+1\right)
}\right) _{\mathrm{M}}=\left\vert \mathbf{P}_{i}^{\left( s\right) }\mathbf{P}%
_{k}^{\left( s\right) }\right\vert _{\mathrm{M}}^{2}-F\left( P_{i}^{\left(
s\right) },P_{k}^{\left( s\right) },P_{i}^{\left( s+1\right) },P_{k}^{\left(
s+1\right) }\right)  \label{a5.2}
\end{equation}%
\begin{equation}
\left\vert \mathbf{P}_{i}^{\left( s\right) }\mathbf{P}_{k}^{\left( s\right)
}\right\vert _{\mathrm{M}}^{2}=\left\vert \mathbf{P}_{i}^{\left( s+1\right) }%
\mathbf{P}_{k}^{\left( s+1\right) }\right\vert _{\mathrm{M}}^{2}-2d\left(
P_{i}^{\left( s\right) },P_{k}^{\left( s\right) }\right) +2d\left(
P_{i}^{\left( s+1\right) },P_{k}^{\left( s+1\right) }\right)  \label{a5.3}
\end{equation}%
where 
\begin{eqnarray}
&&F\left( P_{i}^{\left( s\right) },P_{k}^{\left( s\right) },P_{i}^{\left(
s+1\right) },P_{k}^{\left( s+1\right) }\right)  \notag \\
&=&d\left( P_{i}^{\left( s\right) },P_{k}^{\left( s+1\right) }\right)
+d\left( P_{k}^{\left( s\right) },P_{i}^{\left( s+1\right) }\right) -d\left(
P_{i}^{\left( s\right) },P_{i}^{\left( s+1\right) }\right) -d\left(
P_{k}^{\left( s\right) },P_{k}^{\left( s+1\right) }\right)  \notag \\
&&-2d\left( P_{i}^{\left( s\right) },P_{k}^{\left( s\right) }\right)
\label{a5.5}
\end{eqnarray}%
In relations (\ref{a5.2}) - (\ref{a5.5}) indices $i,k=0,1,...n$ and $s$
takes all integer values. The index "M" means that the scalar products are
calculated in the geometry $\mathcal{G}_{\mathrm{M}}$ of Minkowski. The
quantities $F\left( P_{i}^{\left( s\right) },P_{k}^{\left( s\right)
},P_{i}^{\left( s+1\right) },P_{k}^{\left( s+1\right) }\right) $ describe
some force fields, which appear in the geometry of Minkowski due to mismatch 
$d$. The force fields $F$ have an energy-momentum and an angular momentum,
which provide the conservation law in the space-time geometry of Minkowski.

In general, taking into consideration the force fields $F$, one may
investigate the particle dynamics in the space-time geometry of Minkowski.
However, structure of the force fields $F$ is more complicated (function of
four points), than the structure of the world function $\sigma _{\mathrm{r}}$
(function of two points). It seems to be more reasonable to investigate the
free particle motion in a real (complicated) geometry, than to investigate
the particle motion in unknown force fields of the simple space-time
geometry $\mathcal{G}_{\mathrm{M}}$.

\section{Conclusions}

Thus, in the twentieth century a transition from the nonrelativistic physics
to the relativistic one has been produced only in dynamic equations, but not
in the concept of the particle state. In the nonrelativistic physics the
particle state is described as a point in the phase space. Construction of
the phase space is founded on the continuous space-time geometry (of Newton,
or of Minkowski). Existence of primordially indeterministic particles
(elementary particles) is possible only if the space-time geometry is
multivariant. One cannot construct phase space, because the limit (\ref{a1.1}%
), determining the particle momentum, does not exist in a multivariant
space-time geometry. We are forced to describe the particle state without
limits of the type (\ref{a1.1}).

The relativistic concept of a particle state is realized by means of a
skeleton of a particle. The skeleton consists of several space-time points.
The number of the skeleton points depends on the structure of the elementary
particle. In the simplest case of a pointlike particle its skeleton contains
two points. It is important, that the skeleton describes all characteristics
of the particle, including its mass, charge, momentum and other
characteristics, if they take place, (spin, flower, etc. ). As a result one
obtains a monistic conception, where all fundamental physical phenomena
(including electromagnetic and gravitational interactions) are described in
terms of points of the event space and world functions between them.

Dynamic equations are algebraic equations, formulated in a coordinateless
form. These equations are simpler and more universal, than equations, used
in the conventional theory of elementary particles.

The conventional theory of elementary particles, which uses nonrelativistic
concept of particle state, degenerates into phenomenological conception,
which systematize elementary particles and their reactions. However,
pretenses of the conventional approach to determination of the elementary
particles construction are unfounded, because of inconsistent application of
relativistic concepts.

A physical theory is a relativistic, if the event space (space-time) is
described by one and only one structure: world function $\sigma $. If there
are another geometric structures, for instance, spatial distant $S$, the
physical theory constructed on such a two-structure geometry of the event
space is not relativistic. One can construct a derivative geometric
structure -- time interval $T$ on the basis of the world function $\sigma $
and the space distance $S$. Thereafter one can describe the geometry of the
event space on the basis of two structures: space distance $S$ and time
interval $T$, considering them as independent and ignoring the world
function $\sigma $. Such a two-structure geometry is a Newtonian conception
of the event space.

Such a formulation of the difference between the Newtonian and relativistic
conceptions of the event space does not refer to the way of transformation
of dynamic equations, written in inertial coordinate systems. The invariant
(coordinateless) formulation of the relativity theory looks better, than the
formulation with a reference to the transformation law of dynamic equations.
As we have seen, the reference only to the transformation law and disregard
of the relativistic concept of the particle state may lead to an
inconsistent conception.

\end{document}